\documentstyle[12pt]{article}
\begin{document}
\begin{flushright}
Alberta Thy 18-99\\hep-th/9910189
\end{flushright}
\vspace{1 cm}
\begin{center}
\baselineskip=16pt

{\Large\bf  Generalized Invariants and Quantum Evolution of Open
Fermionic System}

\vskip 2cm {\bf S. P. Kim},$^{a}$\footnote{E-mail:
sangkim@ks.kunsan.ac.kr} {\bf  A. E.
Santana},$^{b}$\footnote{E-mail: santana@fis.ufba.br} {\bf F. C.
Khanna},$^{c}$\footnote{E-mail: khanna@phys.ualberta.ca}
\\ \vskip 0.8cm

${}^{a}$Department of Physics, Kunsan National University, Kunsan
573-701, Korea \\ ${}^{b}$Instituto de Fisica, Universidade
Federal da Bahia, Campus de Ondina, 40210-340, Salvador, Bahia,
Brazil\\ ${}^{c}$Theoretical Physics Institute, Department of
Physics, University of Alberta, Edmonton, Alberta, Canada, T6J 2J1
\\{\rm and} \\ TRIUMF, 4004 Wesbrook Mall, Vancouver, British
Columbia, Canada, V6T 2A3

\vskip 1 cm

\end{center}

\vskip 1 cm

\centerline{\bf ABSTRACT}
\begin{quotation}
Open systems acquire time-dependent coupling constants through
interaction with an external field or environment. We generalize
the Lewis-Riesenfeld invariant theorem to open system of quantum
fields after second quantization. The generalized invariants and
thereby the quantum evolution are found explicitly for
time-dependent quadratic fermionic systems.  The pair production
of fermions is computed and other physical implications are
discussed.
\end{quotation}


\newpage

The dynamical evolution of the Universe in time has to be examined
carefully to understand the phase transitions that are crucial to
the cosmological scenarios. The phase transition may be affecting
topologically unstable structures in the Universe and these may
have important consequences for the inflationary Universe. The
observational aspects like baryon asymmetry may require the
presence of time-dependent gauge fields. The time evolution of the
Universe would require that both scalar and fermionic fields are
time dependent through their interactions with the gravity.
Furthermore one may need to calculate the pair production of
fermions due to the oscillating inflatons. This may provide a
possible mechanism for preheating mechanism for fermionic systems.

All these time-dependent processes in the Universe would be
happening when the system is not in equilibrium and there is no
fixed temperature. We have a system that is time-dependent and in
a non-equilibrium state. Similar time-dependent and rapidly
varying processes in the laboratory, that simulate the conditions
at the early evolutionary stage of the Universe, relate to the
heavy ion reactions at the relativistic heavy ion facility at
Brookhaven. It is intended that in colliding heavy ion  (U on U)
reactions there will be a deconfining phase transition from
nucleonic to a quark-gluon system that exists for a very brief
period of time before making a phase transition back to the
baryonic matter. The time evolution of the formation and
dissolution of the quark-gluon plasma is a highly non-equilibrium
process that requires a careful consideration of the time
dependence of the coupling of both the fermionic and bosonic
systems.

The coupling constants or parameters of a system depend on time
implicitly through the interaction  when it interacts with an
environment. Any system whose coupling constants or parameters
depend on time explicitly can be regarded as a subsystem of a
larger closed system and belongs to an open system.  To describe
properly the quantum states of such open system one may find  the
evolution of the system plus environment and integrate out the
degrees of freedom corresponding to the environment. Or one may
rely on field theoretical methods such as closed time path
integral (for review and references, see \cite{ctpi}) or thermo
field dynamics (for review and references, see \cite{tfd}).

Even though the thermo field dynamics is a real time operator
formalism it has not been applied to a real problem to investigate
the time evolution. Hence we choose instead to use the functional
Schr\"{o}dinger picture approach [3-5]. It is conceivable that
this may provide some help in using the thermo field dynamics.
Even the quantum evolution of a time-dependent system has also
been an outstanding problem in quantum mechanics. A frequently
used method is to find trial wave functions (a variational
approach) with undetermined time-dependent coefficients and to fix
these coefficients by solving the Schr\"{o}dinger equation. Lewis
and Riesenfeld introduced a systematic method to find the
operators that satisfy the quantum Liouville-von Neumann equation
and to solve for the exact quantum states as their eigenstates
\cite{lewis}.  This method has been applied successfully to some
of the time-dependent systems, in particular, the harmonic
oscillator systems.

The Lewis-Riesenfeld invariant theorem has been applied
successfully to study the quantum evolution of scalar fields in an
expanding Friedmann-Robertson-Walker universe [7-12]. A variety of
results were obtained including the vacuum state and the Wightman
functions with respect to the vacuum, thermal equilibrium and
coherent states. However, as suggested above the overall evolution
of the universe would require not only the time dependence of the
scalar field but also that of a fermion field. Since the fermion
field is a four component spinor, it appears natural to use the
Lewis-Riesenfeld invariant theorem to the fermion field taking
proper account of the fact that the spinor components lead to a
set of coupled equations. The usual procedure of writing the
spinor field in terms of creation and annihilation operators is
adopted. The results of quantum evolution of the fermion field are
presented. We give a brief overview of the application to the
scalar field and then a detailed account of the method as applied
to the fermion system is given.

The open system of the quantum field is described by the
functional Schr\"{o}dinger equation [3-5]
\begin{equation}
i \frac{\partial}{\partial t} \vert \Psi (\phi, t) \rangle =
\hat{H} (\phi, -i \frac{\delta}{\delta \phi},t) \vert \Psi (\phi,
t) \rangle,
\end{equation}
where $\phi$ represents a scalar or fermion field. Our stratagem
is to express the Hamiltonian in the second quantization as a sum
of infinite number of finite dimensional quantum systems, which
are decoupled or coupled depending on the type of interactions.
And then we apply the Lewis-Riesenfeld theorem to each finite
dimensional system and find the generalized invariants and thereby
the evolution of the second quantized Hamiltonian.

According to the Lewis-Riesenfeld invariant theorem, the hermitian
invariant operator $\hat{\cal O}(t)$, satisfying the Liouville-von
Neumann equation for a time-dependent Hamiltonian $\hat{H} (t)$
\begin{equation}
i \frac{d}{d t} \hat{\cal O} (t) = i \frac{\partial}{\partial t}
\hat{\cal O} (t) + [ \hat{\cal O} (t), \hat{H} (t) ] = 0,
\label{ln eq}
\end{equation}
has time-dependent eigenstates and time-independent eigenvalues
\begin{equation}
\hat{\cal O} (t) \vert \lambda_n, t \rangle = \lambda_n \vert
\lambda_n, t \rangle. \label{eig eq}
\end{equation}
And the exact quantum evolution of the system
\begin{equation}
i \frac{\partial}{\partial t} \vert \Psi (t) \rangle = \hat{H} (t)
\vert \Psi(t) \rangle, \label{sch eq}
\end{equation}
is given by a superposition of states $\vert \lambda_n, t
\rangle$, up to time-dependent phase factors.

The theorem was originally proved by Lewis and Riesenfeld and
applied to a time-dependent harmonic oscillator \cite{lewis}. We
now give a simple proof of the theorem. We assume that $\hat{\cal
O} (t)$ be a hermitian operator and act on the Hilbert space for
the system $\hat{H}(t)$. When the states are not normalizable, the
Hilbert space should be extended to a rigged one \cite{simon}. The
operator may have a possible degeneracy
\begin{equation}
\hat{\cal O} (t) \vert \lambda^a_n, t \rangle = \lambda_n \vert
\lambda^a_n, t \rangle, \label{deg eig eq}
\end{equation}
where the index $a$ runs over the degenerate eigenstates. One can
apply eq. (\ref{ln eq}) to the eigenstate $\vert \lambda^a_n, t
\rangle$ and differentiate eq. (\ref{eig eq}) to obtain the
following identity
\begin{equation}
\Bigl(i \frac{\partial \lambda_n}{\partial t} \Bigr) \vert
\lambda^a_n, t \rangle - \Bigl(\hat{\cal O} (t) - \lambda_n
\hat{I} \Bigr) \Bigl[ \Bigl( i \frac{\partial}{\partial t} -
\hat{H} (t) \Bigr) \vert \lambda^a_n, t \rangle \Bigr] = 0,
\label{ker eq}
\end{equation}
where $\hat{I}$ denotes the identity operator.  By acting $\langle
\lambda^a_n, t \vert$ on eq. (\ref{ker eq}) and using the fact
that $\hat{\cal O} (t)$ is the hermitian operator and $\langle
\lambda^a_n, t \vert$ belongs to the kernel of the hermitian
operator $\hat{\cal O} (t) - \lambda_n \hat{I}$, one finds that
$\frac{\partial \lambda_n}{\partial t} = 0$, that is, the
eigenvalue is a constant. Hence the first term of eq. (\ref{ker
eq}) vanishes and eq. (\ref{ker eq}) reduces to
\begin{equation}
\Bigl( i \frac{\partial}{\partial t} - \hat{H} (t) \Bigr) \vert
\lambda_n^a, t \rangle =  \sum_{b} \varphi^{ba}_n (t) \vert
\lambda_n^b, t \rangle.
\end{equation}
Assuming the orthonormal basis $\langle \lambda_m^a, t \vert
\lambda_n^b, t \rangle = \delta_{mn} \delta_{ab}$, $\varphi^{ba}_n
(t)$ is determined by acting $\langle \lambda_{n}^{b}, t \vert$ on
the both sides:
\begin{equation}
\varphi^{ba}_n (t) = \langle \lambda_n^b, t \vert \Bigl( i
\frac{\partial}{\partial t} - \hat{H} (t) \Bigr) \vert
\lambda_n^a, t \rangle.
\end{equation}
Since $\hat{\cal O} (t) - \lambda_n \hat{I}$ is a hermitian
operator and $\varphi^{ab}_n (t)$ is an element of the
corresponding hermitian matrix, there exist a set of orthonormal
eigenvectors such that
\begin{equation}
\varphi^{ba}_n (t) C_a (t) = \varphi_n^{\alpha} (t) C_b (t).
\end{equation}
For the new basis for the subspace spanned by eigenvectors $\vert
\lambda_n^a, t \rangle$ for a given $\lambda_n$:
\begin{equation}
\vert \lambda_n^{\alpha}, t \rangle = \sum_{a} C_a (t) \vert
\lambda_n^a, t \rangle,
\end{equation}
one then has
\begin{equation}
\Bigl( i \frac{\partial}{\partial t} - \hat{H} (t) \Bigr) \vert
\lambda_n^{\alpha}, t \rangle = \varphi^{\alpha}_n (t) \vert
\lambda_n^{\alpha}, t \rangle,
\end{equation}
from which follows the equation
\begin{equation}
\varphi^{\alpha}_n (t) = \langle \lambda_n^{\alpha}, t \vert
\Bigl( i \frac{\partial}{\partial t} - \hat{H} (t) \Bigr) \vert
\lambda_n^{\alpha}, t \rangle.
\end{equation}
The quantum state given by
\begin{equation}
\vert \Psi_n, t \rangle = e^{i \int \varphi^{\alpha}_n (t)} \vert
\lambda^{\alpha}_n, t \rangle, \label{ex quan}
\end{equation}
is an exact solution of eq. (\ref{sch eq}). The most general state
is a superposition of states like the ones given by eq. (\ref{ex
quan}).

It should be pointed out that the Lewis-Riesenfeld theorem can
also be applied to infinite-dimensional quantum systems such as
the quantum fields when they are expressed as a sum of finite
dimensional systems. The second quantized Hamiltonian of quantum
fields has such a form.  Another important observation is that the
theorem can also be applied to bosonic or fermionic systems when
the commutators or anticommutators are appropriately used for
boson and fermion operators, respectively. We shall consider a
scalar field with a time-dependent mass and a time-dependent
fermion system with the most general quadratic form as open
systems.

A massive scalar field with a time-dependent mass described by the
Lagrangian density
\begin{equation}
{\cal L} = \frac{1}{2}g^{\mu \nu} \partial_{\mu} \phi
\partial_{\nu} \phi - \frac{m^2 (t)}{2} \phi^2
\end{equation}
has the second quantized Hamiltonian in the standard basis of the
Minkowski spacetime
\begin{equation}
\hat{H} (t) =  \sum_{\alpha} \Omega^{(D)}_{\alpha} (t)
\Bigl(\hat{a}^{\dagger}_{\alpha} \hat{a}_{\alpha} + \frac{1}{2}
\Bigr) + \Omega_{\alpha} (t) \Bigl(\hat{a}^{\dagger 2}_{\alpha} +
\hat{a}^2_{\alpha} \Bigr) \equiv \sum_{\alpha} \hat{H}_{\alpha}
(t). \label{sca ham}
\end{equation}
Here, $\alpha$ denotes collectively the Fourier sine/cosine modes,
and
\begin{equation}
\Omega^{(D)}_{\alpha} (t) = \frac{\omega_{\alpha}^2}{2} +
\frac{1}{2}, ~~ \Omega_{\alpha} (t) = \frac{\omega^2_{\alpha}}{4}
- \frac{1}{4},
\end{equation}
where
\begin{equation}
\omega^2_{\alpha} = m^2 (t) + {\bf k}^2.
\end{equation}
The creation and annihilation operators for each mode satisfy the
usual commutation relation
\begin{equation}
[ \hat{a}_{\alpha}, \hat{a}^{\dagger}_{\beta}] = \delta_{\alpha
\beta}.
\end{equation}
The Lewis-Riesenfeld theorem applied to $\hat{H}_{\alpha} (t)$
gives rise to a pair of invariants [7-12]
\begin{equation}
\hat{A}_{\alpha} (t) = f^{(-)}_{\alpha} (t) \hat{a}_{\alpha} +
f^{(+)}_{\alpha} (t) \hat{a}^{\dagger}_{\alpha}, ~~
\hat{A}^{\dagger}_{\alpha} (t) = {\rm h.c.},
\end{equation}
where
\begin{eqnarray}
&& f^{(-)}_{\alpha} (t) = \frac{1}{\sqrt{2}} \bigl(-i
 \dot{\varphi}^*_{\alpha} (t) + \varphi^*_{\alpha} (t) \bigr),
~~f^{(+)}_{\alpha} (t) = \frac{1}{\sqrt{2}}
 \bigl(-i \dot{\varphi}^*_{\alpha} (t) - \varphi^*_{\alpha} (t) \bigr), \\
&& \ddot{\varphi}_{\alpha} (t) + \omega^2_{\alpha} (t)
\varphi_{\alpha} (t)
 = 0. \label{cl eq}
\end{eqnarray}
It should be remarked that $\varphi_{\bf k} (t)$ satisfies the
classical equation for the ${\bf k}$-mode. The equal time
commutator
\begin{equation}
[ \hat{A}_{\alpha} (t), \hat{A}^{\dagger}_{\beta} (t)] =
\delta_{\alpha \beta},
\end{equation}
is satisfied for all times for a complex solution due to the
wronskian of eq. (\ref{cl eq}):
\begin{equation}
i \Bigl(\dot{\varphi}_{\alpha} (t) \varphi_{\alpha}^* (t) -
\dot{\varphi}^*_{\alpha} \varphi_{\alpha} (t) \Bigr) = 1.
\end{equation}
The number operators and their eigenstates
\begin{equation}
\hat{N}_{\alpha} (t) = \hat{A}^{\dagger}_{\alpha} (t)
\hat{A}_{\alpha} (t), ~~ \hat{N}_{\alpha} \vert n_{\alpha}, t
\rangle = n_{\alpha} \vert n_{\alpha}, t \rangle,
\end{equation}
define the Fock space for each mode. The exact quantum states for
each mode are linear superposition of these number states as given
by eq. (\ref{ex quan}),  and the wave function for the scalar
field  is the product of these  quantum states.

We now generalize the quadratic form of eq. (\ref{sca ham}) to a
fermionic system by defining the Hamiltonian
\begin{eqnarray}
\hat{H} (t) = \sum_{\alpha} \Omega^{(D)}_{\alpha} (t)
\Bigl(\hat{b}^{\dagger}_{\alpha} \hat{b}_{\alpha} -
\hat{d}_{\alpha} \hat{d}^{\dagger}_{\alpha}  \Bigr) +
\Omega^{(--)}_{\alpha} (t) \hat{b}_{\alpha}\hat{d}_{\alpha}
 + \Omega^{(++)}_{\alpha} (t)\hat{b}^{\dagger}_{\alpha}
\hat{d}^{\dagger}_{\alpha} \nonumber\\ + \Omega^{(-+)}_{\alpha}
(t)\hat{b}_{\alpha} \hat{d}^{\dagger}_{\alpha} +
\Omega^{(+-)}_{\alpha} (t) \hat{b}^{\dagger}_{\alpha}
\hat{d}_{\alpha}.   \label{gen ham}
\end{eqnarray}
Here, $\alpha$ denotes a collective notation for the momentum
${\bf k}$ and spinor index $r$, and $\sum_{\alpha} = \int d^3 {\bf
k}  \sum_{r =1, 2}$, and the particle and antiparticle creation
and annihilation operators satisfy the anticommutators,
respectively,
\begin{equation}
\{\hat{b}_{\alpha}, \hat{b}^{\dagger}_{\beta} \} = \delta_{\alpha
\beta},~~ \{\hat{d}_{\alpha}, \hat{d}^{\dagger}_{\beta} \} =
\delta_{\alpha \beta},
\end{equation}
and all the other anticommutators vanish. So the Hamiltonian given
by eq. (\ref{gen ham}) describes a system bilinear in the fermion
fields. The unitarity of quantum evolution requires the
hermiticity of the Hamiltonian which leads to the conditions
\begin{equation}
\Omega^{(D)}_{\alpha} (t) = \Omega^{(D)*}_{\alpha} (t),~~
\Omega^{(--)}_{\alpha} (t) = - \Omega^{(++)*}_{\alpha}
(t),~~\Omega^{(-+)}_{\alpha} (t) = - \Omega^{(+-)*}_{\alpha} (t).
\label{dec eq}
\end{equation}

It is found that an invariant of the form
\begin{equation}
\hat{\cal O}_{\alpha} (t) = f^{(-)}_{\alpha} (t) \hat{b}_{\alpha}
+ f^{(+)}_{\alpha} (t) \hat{b}^{\dagger}_{\alpha} +
g^{(-)}_{\alpha} (t) \hat{d}_{\alpha} + g^{(+)}_{\alpha} (t)
\hat{d}^{\dagger}_{\alpha},\label{inv op}
\end{equation}
satisfies eq. (\ref{ln eq}) provided that
\begin{eqnarray}
i \frac{\partial}{\partial t} f^{(-)}_{\alpha} (t) +
\Omega_{\alpha}^{(D)} (t) f^{(-)}_{\alpha} (t) -
\Omega^{(-+)}_{\alpha} (t)
 g^{(-)}_{\alpha}(t) - \Omega^{(--)}_{\alpha}(t) g^{(+)}_{\alpha} (t) = 0, \nonumber\\
i \frac{\partial}{\partial t} f^{(+)}_{\alpha} (t) -
\Omega_{\alpha}^{(D)} (t) f^{(+)}_{\alpha} (t) -
\Omega^{(++)}_{\alpha} (t) g^{(-)}_{\alpha} (t) -
\Omega^{(+-)}_{\alpha} (t) g^{(+)}_{\alpha} (t) = 0, \nonumber\\ i
\frac{\partial}{\partial t} g^{(-)}_{\alpha} (t) +
\Omega_{\alpha}^{(D)} (t) g^{(-)}_{\alpha} (t) +
\Omega^{(+-)}_{\alpha} (t) f^{(-)}_{\alpha} (t) +
\Omega^{(--)}_{\alpha}(t) f^{(+)}_{\alpha} (t) = 0, \nonumber\\ i
\frac{\partial}{\partial t} g^{(+)}_{\alpha} (t) -
\Omega_{\alpha}^{(D)} (t) g^{(+)}_{\alpha} (t) +
\Omega^{(++)}_{\alpha} (t) f^{(-)}_{\alpha} (t) +
\Omega^{(-+)}_{\alpha} (t) f^{(+)}_{\alpha} (t) = 0. \label{fer
eq}
\end{eqnarray}
By introducing two column vectors
\begin{equation}
U_{\alpha} (t) = \frac{1}{\sqrt{2}}\pmatrix{f^{(-)}_{\alpha} (t) +
f^{(+)}_{\alpha}(t) \cr\\ f^{(-)}_{\alpha} (t) - f^{(+)}_{\alpha}
(t)},~~ V_{\alpha} (t) =
\frac{1}{\sqrt{2}}\pmatrix{g^{(-)}_{\alpha} (t) + g^{(+)}_{\alpha}
(t) \cr\\ g^{(-)}_{\alpha} (t) -  g^{(+)}_{\alpha}(t)},
\end{equation}
we can write eq. (\ref{fer eq}) as two vector equations
\begin{eqnarray}
i \frac{\partial}{\partial t} U_{\alpha} (t) +
\Omega^{(D)}_{\alpha} (t) \sigma_1 U_{\alpha} (t) + M_{U_{\alpha}}
(t) V_{\alpha} (t) = 0, \\ i \frac{\partial}{\partial t}
V_{\alpha} (t) + \Omega^{(D)}_{\alpha} (t) \sigma_1 V_{\alpha} (t)
+ M_{V_{\alpha}} (t) U_{\alpha} (t) = 0.\label{vec eq}
\end{eqnarray}
Here, $M_{U_{\alpha}} (t)$ and $M_{V_{\alpha}} (t)$ denote the
mixing matrices between $U_{\alpha} (t)$ and $V_{\alpha} (t)$:
\begin{eqnarray}
M_{U_{\alpha}} (t) =  \Delta^{(0)}_{\alpha} (t) I +
\Delta^{(1)}_{\alpha} (t) \sigma_{1}+\Delta^{(2)}_{\alpha} (t)
\sigma_{2}+\Delta^{(3)}_{\alpha} (t) \sigma_{3}, \nonumber\\
M_{V_{\alpha}} (t) =  - \Delta^{(0)}_{\alpha} (t) I +
\Delta^{(1)}_{\alpha} (t) \sigma_{1} - \Delta^{(2)}_{\alpha} (t)
\sigma_{2} - \Delta^{(3)}_{\alpha} (t) \sigma_{3},
\end{eqnarray}
where $\sigma_{i}$ are Pauli spin matrices and
\begin{eqnarray}
\Delta^{(0)}_{\alpha} (t) = - \frac{1}{2} \Bigl(
\Omega^{(-+)}_{\alpha} (t)
 - \Omega^{(-+)*}_{\alpha} (t) \Bigr),~~
\Delta^{(1)}_{\alpha}(t) = - \frac{1}{2} \Bigl(
\Omega^{(-+)}_{\alpha} (t) + \Omega^{(-+)*}_{\alpha} (t) \Bigr),
\nonumber\\ \Delta^{(2)}_{\alpha} (t) = \frac{i}{2}
\Bigl(\Omega^{(--)}_{\alpha} (t) + \Omega^{(--)*}_{\alpha} (t)
\Bigr),~~ \Delta^{(3)}_{\alpha} (t) = \frac{1}{2}
\Bigl(\Omega^{(++)}_{\alpha} (t) - \Omega^{(++)*}_{\alpha} (t)
\Bigr).
\end{eqnarray}
By eliminating either $U_{\alpha}(t)$ or $V_{\alpha}(t)$ in eq.
(\ref{vec eq}) we obtain  the second order differential equations.
So we get two types of annihilation operators
\begin{eqnarray}
\hat{B}_{\alpha} (t) = f^{(B-)}_{\alpha} (t) \hat{b}_{\alpha} +
f^{(B+)}_{\alpha} (t) \hat{b}^{\dagger}_{\alpha} +
g^{(B-)}_{\alpha} (t) \hat{d}_{\alpha} + g^{(B+)}_{\alpha} (t)
\hat{d}^{\dagger}_{\alpha}, \nonumber\\ \hat{D}_{\alpha} (t) =
f^{(D-)}_{\alpha} (t) \hat{b}_{\alpha} + f^{(D+)}_{\alpha} (t)
\hat{b}^{\dagger}_{\alpha} + g^{(D-)}_{\alpha} (t)
\hat{d}_{\alpha} + g^{(D+)}_{\alpha} (t)
\hat{d}^{\dagger}_{\alpha}, \label{fer inv}
\end{eqnarray}
where $f^{(B-)}_{\alpha}(t)$ and $g^{(D-)}_{\alpha}(t)$ are chosen
the (adiabatic) positive frequency solutions.

One can make $\hat{B}_{\alpha}(t), \hat{D}_{\alpha} (t)$ and their
hermitian conjugates the annihilation and creation operators that
satisfy the anticommutators at each time
\begin{equation}
\{ \hat{B}_{\alpha}(t), \hat{B}^{\dagger}_{\beta}(t) \} =
\delta_{\alpha \beta},~~ \{ \hat{D}_{\alpha}(t),
\hat{D}^{\dagger}_{\beta}(t) \} = \delta_{\alpha \beta}.
\end{equation}
These lead to the consistency condition for the coefficient
functions
\begin{eqnarray}
f^{(B-)*}_{\alpha} (t)  f^{(B-)}_{\alpha} (t) + f^{(B+)*}_{\alpha}
(t) f^{(B+)}_{\alpha} (t)+ g^{(B-)*}_{\alpha} (t)
g^{(B-)}_{\alpha} (t) + g^{(B+)*}_{\alpha} (t)  g^{(B+)}_{\alpha}
(t) = 1, \nonumber\\ f^{(D-)*}_{\alpha} (t)  f^{(D-)}_{\alpha} (t)
+ f^{(D+)*}_{\alpha} (t) f^{(D+)}_{\alpha} (t)+ g^{(D-)*}_{\alpha}
(t)  g^{(D-)}_{\alpha} (t) + g^{(D+)*}_{\alpha} (t)
g^{(D+)}_{\alpha} (t) = 1. \label{anti con}
\end{eqnarray}
The consistency of anticommutators for all times is guaranteed by
\begin{eqnarray}
\frac{\partial}{\partial t} \Bigl(U^{(B)\dagger} (t) U^{(B)} (t) +
V^{(B)\dagger} (t) V^{(B)} (t) \Bigr) = 0, \nonumber\\
\frac{\partial}{\partial t} \Bigl(U^{(D)\dagger} (t) U^{(D)} (t) +
V^{(D)\dagger} (t) V^{(B)} (t) \Bigr) = 0,
\end{eqnarray}
where $M^{\dagger}_{U_{\alpha}} (t) = M_{V_{\alpha}} (t)$ is used.
The other anticommutators
\begin{eqnarray}
\{ \hat{B}_{\alpha}(t), \hat{B}_{\beta}(t)\} = \{
\hat{B}_{\alpha}^{\dagger}(t), \hat{B}_{\beta}^{\dagger}(t)\} =
0,\nonumber\\ \{ \hat{D}_{\alpha}(t), \hat{D}_{\beta}(t)\} = \{
\hat{D}_{\alpha}^{\dagger}(t), \hat{D}_{\beta}^{\dagger}(t)\} = 0,
\end{eqnarray}
are consistently satisfied for all times due to
\begin{eqnarray}
\frac{\partial}{\partial t} \Bigl(U^{(B)\dagger} (t) \sigma_{3}
U^{(B)} (t) + V^{(B)\dagger} (t) \sigma_{3} V^{(B)} (t) \Bigr) =
0, \nonumber\\ \frac{\partial}{\partial t} \Bigl(U^{(D)\dagger}
(t) \sigma_{3} U^{(D)} (t) + V^{(D)\dagger} (t) \sigma_{3} V^{(B)}
(t) \Bigr) = 0,
\end{eqnarray}
where  $\sigma_3 M^{T}_{U_{\alpha}} (t) \sigma_3 =  -
M_{V_{\alpha}}(t)$ is used. Still other anticommutators
\begin{eqnarray}
\{ \hat{B}_{\alpha}(t), \hat{D}_{\beta}(t)\} = \{
\hat{B}_{\alpha}^{\dagger}(t), \hat{D}_{\beta}^{\dagger}(t)\} =
0,\nonumber\\ \{ \hat{B}_{\alpha}(t), \hat{D}_{\beta}^{\dagger}
(t)\} = \{ \hat{B}_{\alpha}^{\dagger}(t), \hat{D}_{\beta} (t)\} =
0,
\end{eqnarray}
are also satisfied for all times due to
\begin{eqnarray}
\frac{\partial}{\partial t} \Bigl(U^{(B)T} (t) \sigma_{3} U^{(D)}
(t) + V^{(B)T} (t) \sigma_{3} V^{(D)} (t) \Bigr) = 0, \nonumber\\
\frac{\partial}{\partial t} \Bigl(U^{(B)\dagger} (t) U^{(D)} (t) +
V^{(B)\dagger} (t) V^{(D)}_{\alpha} (t) \Bigr) = 0.
\end{eqnarray}

Since the $\hat{B}_{\alpha}(t), \hat{B}^{\dagger}_{\alpha}(t)$ and
$\hat{D}_{\alpha}(t), \hat{D}^{\dagger}_{\alpha}(t)$ share the
same properties of the annihilation and creation operators for
time-independent system, they will be used as those for
time-dependent system. So the number operators are defined by
\begin{equation}
\hat{N}^{(B)}_{\alpha} (t) =  \hat{B}^{\dagger}_{\alpha} (t)
\hat{B}_{\alpha} (t), ~~\hat{N}^{(D)}_{\alpha} (t) =
\hat{D}^{\dagger}_{\alpha} (t) \hat{D}_{\alpha} (t).
\end{equation}
The eigenstates for each mode are the number states of
$\hat{N}^{(B)}_{\alpha}(t)$  or $\hat{N}^{(D)}_{\alpha}(t)$. For
instance, the zero-particle state for $\hat{N}^{(B)}_{\alpha}(t)$
is given by
\begin{equation}
\vert 0^{(B)}_{\alpha}, t \rangle = c^{(B)}_{\alpha 0} (t) \vert
0^{(b)}_{\alpha}, 0^{(d)}_{\alpha} \rangle + c^{(B)}_{\alpha 1}
(t) \vert  0^{(b)}_{\alpha} ,  1^{(d)}_{\alpha} \rangle +
e^{(B)}_{\alpha 0} (t) \vert  1^{(b)}_{\alpha},  0^{(d)}_{\alpha}
\rangle + e^{(B)}_{\alpha 1} (t) \vert  1^{(b)}_{\alpha},
1^{(d)}_{\alpha} \rangle, \label{zero st}
\end{equation}
where $\vert n'^{(b)}, n^{(d)} \rangle$ are the number states with
respect to $\hat{b}^{\dagger}_{\alpha} \hat{b}_{\alpha}$ and
$\hat{d}^{\dagger}_{\alpha}  \hat{d}_{\alpha}$, respectively,  and
\begin{equation}
f^{(B -)}_{\alpha} (t) f^{(B +)}_{\alpha} (t) - g^{(B -)}_{\alpha}
(t) g^{(B +)}_{\alpha} (t) = 0.
\end{equation}
One-particle state is constructed by applying the operator
$\hat{B}^{\dagger}_{\alpha} (t)$ to the zero-particle state (eq.
(\ref{zero st})). The number states of the antiparticle can be
similarly constructed by using the number operator
$\hat{N}^{(D)}_{\alpha} (t)$. These states in general depend on
time, but their eigenvalues (occupation numbers) do not depend on
time as expected according to the Lewis-Riesenfeld theorem. The
vacuum state is the one that is annihilated by all the
annihilation operators $\hat{B}_{\alpha} (t)$ and
$\hat{D}_{\alpha} (t)$:
\begin{equation}
\hat{B}_{\alpha} (t) \vert 0, 0, t \rangle = \hat{D}_{\alpha} (t)
\vert 0, 0, t \rangle = 0. \label{vac}
\end{equation}
When the interaction is turned on for a finite period of time or
the system evolves from an asymptotic region where all the
$\Omega_{\alpha}$ are constants and eq. (\ref{vac}) coincides with
the Minkowski vacuum, the pair production of fermions from the
initial vacuum is given by
\begin{equation}
\langle 0, 0 \vert \sum_{\alpha} \Bigl(\hat{N}^{(B)}_{\alpha} (t)
+ \hat{N}^{(D)}_{\alpha} (t) \Bigr) \vert 0, 0 \rangle =
\sum_{\alpha} \Bigl( f^{(B+)*}_{\alpha} (t) f^{(B+)}_{\alpha} +
g^{(D+)*}_{\alpha} (t)  g^{(D+)}_{\alpha} \Bigr). \label{pair
prod}
\end{equation}

Let us consider some particular examples of using the Hamiltonian
given by eq. (\ref{gen ham}).

\noindent {\it Dirac Oscillator.} It is the case where
$\Omega_{\alpha}^{(D)} \neq 0$ and $\Omega^{(\pm, \pm)}_{\alpha} =
0$. Then the first order equations (eq. (\ref{fer eq})) can be
easily integrated as $\hat{B}_{\alpha} (t) = e^{i \int
\Omega_{\alpha}^{(D)}} \hat{b}_{\alpha}$ and $\hat{D}_{\alpha} (t)
= e^{i \int \Omega_{\alpha}^{(D)}} \hat{d}_{\alpha}$ and their
hermitian conjugates. So the generalized invariants become
$\hat{B}^{\dagger}_{\alpha} (t) \hat{B}_{\alpha} (t) =
\hat{b}^{\dagger}_{\alpha} \hat{b}_{\alpha}$ and $
\hat{D}^{\dagger}_{\alpha} (t) \hat{D}_{\alpha} (t) =
\hat{d}^{\dagger}_{\alpha} \hat{d}_{\alpha}$. Their number states
are not only the eigenstates of the generalized invariants but
also those that diagonalize the Hamiltonian.

\noindent{\it Fermion Scalar Field Coupling.} The case where
$\Omega^{(-+)}_{\alpha} = \Omega^{(+-)}_{\alpha} = 0$,
$\Omega^{(D)}_{\alpha} =  k_0 + \frac{m g \phi(t)}{2 k_0}$ and
$\Omega^{(--)}_{\alpha} = - \Omega^{(++)}_{\alpha} = \frac{m g
\phi(t)}{2 k_0}$ leads to  the Hamiltonian
\begin{eqnarray}
\hat{H} (t) = \int d^3 {\bf k} \Biggl[ \Bigl( k_0 + \frac{m g
\phi(t)}{2 k_0} \Bigr) \sum_{r =1, 2} \Bigl\{\hat{b}^{\dagger}_{r}
({\bf k}) \hat{b}_{r} ({\bf k}) - \hat{d}_{r} ({\bf k})
\hat{d}_{r}^{\dagger} ({\bf k}) \Bigr\}  \nonumber\\ +  \frac{m g
\phi(t)}{2 k_0} \sum_{r =1, 2} \Bigl\{\hat{b}^{\dagger}_{r} ({\bf
k}) \hat{d}_{r}^{\dagger} ({\bf k}) + \hat{d}_{r} ({\bf k})
\hat{b}_{r} ({\bf k}) \Bigr\} \Biggr]. \label{yuk ham}
\end{eqnarray}
This describes a fermion coupled to a uniform scalar field through
the Yukawa coupling
\begin{equation}
{\cal L} = \frac{1}{2} \Bigl( \overline{\psi} i \gamma^{\mu}
(\partial_{\mu} \psi) - (\partial_{\mu} \overline{\psi}) i
\gamma^{\mu} \psi \Bigr) - m \overline{\psi} \psi - \lambda \phi
(t) \overline{\psi} \psi,
\end{equation}
where $\lambda$ is a coupling constant. The coefficient functions
are found to satisfy
\begin{eqnarray}
\frac{\partial^2}{\partial t^2} && \Bigl(f^{(-)}_{\alpha} +
g^{(+)} \Bigr) - \frac{1}{\Omega^{(D)}_{\alpha}} \frac{\partial
\Omega^{(D)}_{\alpha}}{\partial t} \Bigl(f^{(-)}_{\alpha} +
g^{(+)} \Bigr) \nonumber\\ && + \Bigl\{\Omega^{(D)2}_{\alpha} +
\Omega^{(--)2}_{\alpha} + i \Omega^{(D)}_{\alpha}
\frac{\partial}{\partial t}
\Bigl(\frac{\Omega^{(--)}_{\alpha}}{\Omega^{(D)}_{\alpha}}\Bigr)
\Bigr\} \Bigl(f^{(-)}_{\alpha} + g^{(+)} \Bigr) = 0, \nonumber\\
\frac{\partial^2}{\partial t^2} && \Bigl(f^{(-)}_{\alpha} -
g^{(+)} \Bigr) - \frac{1}{\Omega^{(D)}_{\alpha}} \frac{\partial
\Omega^{(D)}_{\alpha}}{\partial t} \Bigl(f^{(-)}_{\alpha} -
g^{(+)} \Bigr) \nonumber\\ && + \Bigl\{\Omega^{(D)2}_{\alpha} +
\Omega^{(--)2}_{\alpha} - i \Omega^{(D)}_{\alpha}
\frac{\partial}{\partial t}
\Bigl(\frac{\Omega^{(--)}_{\alpha}}{\Omega^{(D)}_{\alpha}}\Bigr)
\Bigr\} \Bigl(f^{(-)}_{\alpha} - g^{(+)} \Bigr) = 0.
\end{eqnarray}
The positive solution $f^{(-)}_{\alpha} (t)$ and the negative one
$g^{(+)} (t)$ with the condition (\ref{anti con}) give the correct
pair production rate. Interestingly enough the free fermion in the
Friedmann-Robertson-Walker universe has the same form of
Hamiltonian as eq. (\ref{yuk ham}) \cite{halliwell}.

A few comments are in order. The generalized invariants in eq.
(\ref{fer inv}) for the Hamiltonian given by eq. (\ref{gen ham})
are more general than those for eq. (\ref{yuk ham}), since
$\hat{B}_{\alpha}(t)$ involves only $\hat{b}_{\alpha}$ and
$\hat{d}^{\dagger}_{\alpha}$ and $\hat{D}_{\alpha}(t)$ involve
only $\hat{b}^{\dagger}_{\alpha}$ and $\hat{d}_{\alpha}$. When one
has cross terms like $\hat{b}_{\alpha} \hat{d}^{\dagger}_{\alpha}$
and $\hat{b}^{\dagger}_{\alpha} \hat{d}_{\alpha}$, then eq.
(\ref{fer inv}) is the unique choice compatible with the algebra.
Furthermore, eq. (\ref{pair prod}), when restricted to eq.
(\ref{yuk ham}) gives the same result for the pair production as
other quantum field theories. One may also extend the analysis to
the fermion interacting with an electromagnetic field by
introducing a Hamiltonian that is similar in  form to eq.
(\ref{gen ham}) but involves Dirac matrices $\gamma_{\mu}$ in a
coupling with the time dependent coefficient $\Omega_{\alpha}
(t)$.

One of the important issues not treated in this paper is the
non-equilibrium aspect of open system of quantum fields. This can
be done by introducing a density operator  of the form $
\hat{\rho}_{\alpha} (t) = e^{- \beta(\hat{N}^{(B)}_{\alpha} (t) +
\hat{N}^{(D)}_{\alpha} (t))}$, which satisfies the Liouville-von
Neumann equation by its construction, which will be addressed in a
future publication.  One of the applications of the result of this
paper is the evolution of fermions in an expanding Universe, where
fermions gain time-dependence through the interaction with the
gravity. The other is to calculate the pair production rate of
fermions due to the oscillating inflaton and to provide a
preheating mechanism for the fermionic system. Finally the time
evolution of the quark-gluon plasma in a relativistic heavy ion
collision would simulate the conditions present at the early
moments of the Universe and would thus provide us with an
experimental view of the evolution of the Universe.

After the submission of this paper we were informed of Ref.
\cite{finelli}, where linear invariants have also been introduced
for fermionic systems. However, we have explicitly shown the
auxiliary equations for invariants, proved the consistency of the
method and constructed the Hilbert space of particle and
antiparticle states for general fermionic systems.

\vspace{1.5ex}
\begin{flushleft}
{\large\bf Acknowledgements}
\end{flushleft}

SPK and AES would like to express their appreciation for warm
hospitality of the Theoretical Physics Institute, University of
Alberta where this paper was completed. Also we would like to
thank D.N. Page and Y. Takahashi for many useful discussions, and
F. Finelli for sending Ref. \cite{finelli}. FCK was supported by
NSERC of Canada, SPK by the Korea Research Foundation under Grant
No. 1998-001-D00364, and AES by the CNPq of Brazil.

\end{document}